\documentclass[a4paper, 10pt, twocolumn]{article}

\usepackage[utf8]{inputenc}         % Schriftkodierung dieser Datei
\usepackage[english]{babel}         % für englische Dokumente

\usepackage{graphicx}               % optional für Grafiken
\usepackage{tabularx}               % optional für Tabellen
\usepackage{multirow}               % optional für Tabellen
\usepackage{url}                    % optional für Internet Links

\usepackage[small,bf]{caption}     % bitte für Bildunterschriften verwenden
\usepackage{parskip}
\usepackage{titlesec}
\usepackage{amsmath}                % optional für Formeln

\usepackage{hyperref}  % not originally in template

\titleformat{\section}{\normalfont\large\bfseries}{\thesection}{}{}
\titleformat{\subsection}{\normalfont\large\bfseries}{\thesection}{}{}
\titleformat{\paragraph}{\normalfont\bfseries}{\theparagraph}{}{}
\titlespacing{\section}{0pt}{6pt}{-1pt}
\titlespacing{\subsection}{0pt}{3pt}{-1pt}
\titlespacing{\paragraph}{0pt}{3pt}{-1pt}

\newcolumntype{Y}{>{\centering\arraybackslash}X}    %für Tabellen mit tabularx

\begin{document}

\date{}                                         % kein Datum auf 1. Seite

\title{\vspace{-8mm}\textbf{\large
Visualising and Explaining Deep Learning Models for Speech Quality Prediction}}

% Hier die Namen und Daten der beteiligten Autoren eintragen
\author{
Henrik Tilkorn, Gabriel Mittag$^1$, Sebastian Möller$^{1,2}$ \\
$^1$ \emph{\small Quality and Usability Lab, TU Berlin
}\\
$^2$ \emph{\small Language Technology, DFKI Berlin
}\\
}
\maketitle

\thispagestyle{empty}          
\section*{Abstract}
Estimating quality of transmitted speech is known to be a non-trivial task. While traditionally, test participants are asked to rate the quality of samples; nowadays, automated methods are available. These methods can be divided into: 1) intrusive models, which use both, the original and the degraded signals, and 2) non-intrusive models, which only require the degraded signal. Recently, non-intrusive models based on neural networks showed to outperform signal processing based models. However, the advantages of deep learning based models come with the cost of being more challenging to interpret. To get more insight into the prediction models the non-intrusive speech quality prediction model NISQA is analyzed in this paper. NISQA is composed of a convolutional neural network (CNN) and a recurrent neural network (RNN). The task of the CNN is to compute relevant features for the speech quality prediction on a frame level, while the RNN models time-dependencies between the individual speech frames. Different explanation algorithms are used to understand the automatically learned features of the CNN. In this way, several interpretable features could be identified, such as the sensitivity to noise or strong interruptions. On the other hand, it was found that multiple features carry redundant information.

\section*{Introduction}
A common method to rate the quality of a speech sample is the evaluation by test participants using a 5-point scale, where one corresponds to the lowest and five to the highest quality. These ratings are averaged over all participants to acquire the mean opinion score (MOS). This procedure is inconvenient since it is very time and cost-intensive.\\
To overcome these disadvantages, multiple automated methods have been developed and applied successfully in practice. With the bandwidth extension to super-wideband, these methods additionally had to adapt to this more challenging setting. The speech quality assessment model NISQA proposed by Mittag and Möller \cite{mittag_a} is able to estimate the MOS value of super-wideband speech transmission in a non-intrusive fashion. 

\section*{Methodology}
NISQA receives as inputs the degraded signals in form of a spectrogram. This spectrogram is divided into multiple shorter segments before being processed. To understand which features of the input have a high influence on the decision making several different techniques were used.

\subsubsection*{Explanation approaches}
Occlusion Sensitivity is a straightforward approach, where small regions of the input are masked. If the classification score, namely the estimated speech quality, for the occluded input is lower than the score of the original, the masked part has a positive influence on the overall quality. On the other hand is an increasing classification score a sign of a hidden negative influence. To get a complete overview of all the influences of different regions the mask is gradually moved over the input. To use Occlusion Sensitivity a few design choices have to be made. The size and the shape of the mask have to be determined and secondly, the stride, which is how far the mask gets shifted against its previous location, has to be set. To choose a proper size and shape it is beneficial to include prior knowledge about the inputs. In this case, spectrograms were scrutinized and therefore it is known that the width-axis corresponds to time and the height-axis displays the presence of certain frequencies. In general, it is possible that disruptions that lead to a decrease of the speech quality occur only at a particular time and therefore would be detected by a vertically oriented mask, or occur only on a certain frequency and then could be detected by a horizontal mask. The size of the stride mainly controls how fine-grained the overall result is. If the stride is chosen smaller than the mask size the masks of consecutive inputs are overlapping and an averaging step is needed to get the interpretable results. Despite being an intuitive approach, Occlusion Sensitivity has a shortcoming that is hard to avoid. To get a detailed enough explanation, the size and the stride have to be chosen relatively small, which leads, in combination with rather large input images, to too many masked versions to be evaluated. Since disturbances can occur in different shapes it is also necessary to use variously shaped and oriented masks which again leads to many evaluations.\\
While for Occlusion Sensitivity no knowledge about the actual model was needed, the following methods are explicitly designed to explain the decisions of neural networks.\\
DeepLIFT (DL) \cite{shrikumar} is a backpropagation-based approach to analyze the decisions of neural networks. To estimate the relevance scores the model receives the sample of interest and a neutral reference input. The reference input has to be chosen in such a way that the model can't detect any features. In the context of speech quality assessment, an empty spectrogram will suffice, since it has no features that have a harmful or beneficial influence on the speech quality. The model will produce a prediction for the sample and the neutral reference each. The difference between these two predictions is called difference-from-reference. To obtain the actual explanation, the difference-from-reference is propagated back through the model using several propagation rules. \\
Another explanation algorithm is Integrated Gradients (IG) proposed by Sundararajan et al. in 2017 \cite{sundararajan}. Many gradient-based explanations suffer from several shortcomings, some of which are discussed in the section \hyperref[section:challenges]{'Challenges of explanations'}. To overcome these shortcomings IG was designed following the axioms of sensitivity and implementation invariance, which are also described in the following section. Similar to DL, IG uses a neutral reference input called baseline. The model can be understood as a function mapping from the input to an output space. The explanation is then obtained by calculating the path-integral of the gradient along the straight path between the input sample and the baseline. Since calculating the gradient directly is often difficult or not possible a Riemann approximation is used. Further details can be found in the the original work.\\
While DL and IG aim to create explanations on the input level, the Conductance approach takes a closer look at the hidden units of the model. Conductance \cite{dhamdhere}, proposed by Dhamdhere et al., is an extension to IG and can be understood as the flow of IG values through a hidden unit. The goal is to identify which patterns of the input cause an activation of particular units.
\subsubsection*{Challenges of explanations} \label{section:challenges}
It is often difficult to distinguish between not understandable reasoning of a complex model and a bad explanation method. Therefore, it is helpful to assure some properties by design and anticipate well known problems from the beginning. Some of these problems and properties will be elaborated hereafter.\\
The thresholding problem can occur if the gradient is used as an indicator for relevance. The gradient of some functions like ReLU is not continuous around their thresholding point. Consequently, the assigned relevance for two arbitrary close points can vary significantly. \\
The saturation problem can arise if a function depends on multiple inputs. If one of these inputs is sufficiently large (or small) it can saturate the function, i.e. the function reaches its maximal output value by only having this one input. A backpropagation-based explanation method may now allocate all the relevance to this dominant input and disregard the other inputs, despite their potential influences. \\ Sensitivity is one of the axioms used to design Integrated Gradients. A model is sensitive if (a) every difference between input and baseline that leads to a change in classification is assigned a non-zero relevance and (b) every variable that doesn't influence the classification has a relevance score of zero.\\
The second axiom used for IG is implementation invariance. It states that functional equivalent models should yield the same explanation. Two models are functionally equivalent if they produce the same output given the same input, independent of the underlying implementation.\\
The completeness criterion assures that relevance scores attributed to the input sum up to exactly the difference from reference in the output layer. In other words, the amount of relevance is calculated in the forward pass and on the backward pass no relevance is disregarded or artificially added.\\
The advanced explanation methods introduced before account for these problems and fulfill these properties, except for DL not being implementation invariant.    

\section*{Experiments}
The experiments were conducted on the ``SwissQual P.OLQA SWB 503'' dataset \cite{P863}. It consists of 216 samples with 54 conditions. The samples from each condition share a specific kind of disturbance. All samples belong to a test set and therefore weren't used during training.\\
Occlusion Sensitivity can be applied straightforwardly. The spectrograms from the dataset are images with a height of 48px and a width of 1300px. The masks were applied before the spectrograms were divided into smaller segments. To discover as many reasons for decreased speech quality as possible masks with different shapes and orientations were used. To detect very local events a square mask of size 24x24px was used. To find temporal or frequency-related disturbances horizontal masks with different heights (2,4,6,8px) and vertical filters with a width of 13px were used respectively. For all these experiments a stride identical to the mask size was used. The model then estimates a MOS value for every masked version. To create the actual explanation, it is necessary to keep track of the positions of the mask. Each region can then be colored according to the influence it had while being masked. \\
The remaining methods, namely DL, IG and Conductance, are all implemented in the Captum library \cite{kokhlikyan2020captum}. Captum is a collection of many different explanation algorithms designed for PyTorch models. These implementations however aim to evaluate CNN. Since NISQA is consisting of a CNN and a LSTM, only the CNN part, which is responsible for the feature extraction, can be examined.\\
NISQA's CNN receives one segment of the spectrogram at a time and has 20 output channels. To keep the number of visual explanations in a processable range, two different approaches were used. First, only 30 segments were examined, but with respect to each of the 20 output channels. Secondly, every 1300 segments of a sample were evaluated, but only with respect to one output channel at the time. Since these segments are highly overlapping an average was calculated to get the final explanation.

\section*{Results}
Occlusion Sensitivity validates two expected behaviors of NISQA. First, only non-empty regions affect the estimation of the speech quality. If this was not true either NISQA or the implementation of Occlusion Sensitivity would be flawed. The second immediate result is that the prior speech quality of a sample affects how big the influence of masking some parts is. A sample with very high quality, for example, a MOS value over 4, is likely to be negatively influenced by the mask. On the other hand, the quality of a sample with low quality most likely will be improved by masking. It could also be shown that hiding noise can improve the estimated quality in some cases. This is not a contradiction considering how noise influences the perceived quality. It may be less disturbing if there is a constant background noise instead of a noisy part sharply interrupted by a short moment of silence. Nevertheless, a complete lack of noise will lead to a higher quality in general. It is also worth mentioning that masking strong disturbances, which are clearly visible on the spectrogram, does not necessarily lead to a quality improvement. Again it is likely that the introduced interruption has a stronger negative effect than the previously present perturbations. It remains to emphasize that Occlusion Sensitivity is a rather simple approach and therefore the results shouldn't be over-interpreted. While it is useful to validate some general assumptions a more fine-grained approach should be used for further insights.

\begin{figure}[h!]
    \centering
    \includegraphics[width=0.2\textwidth]{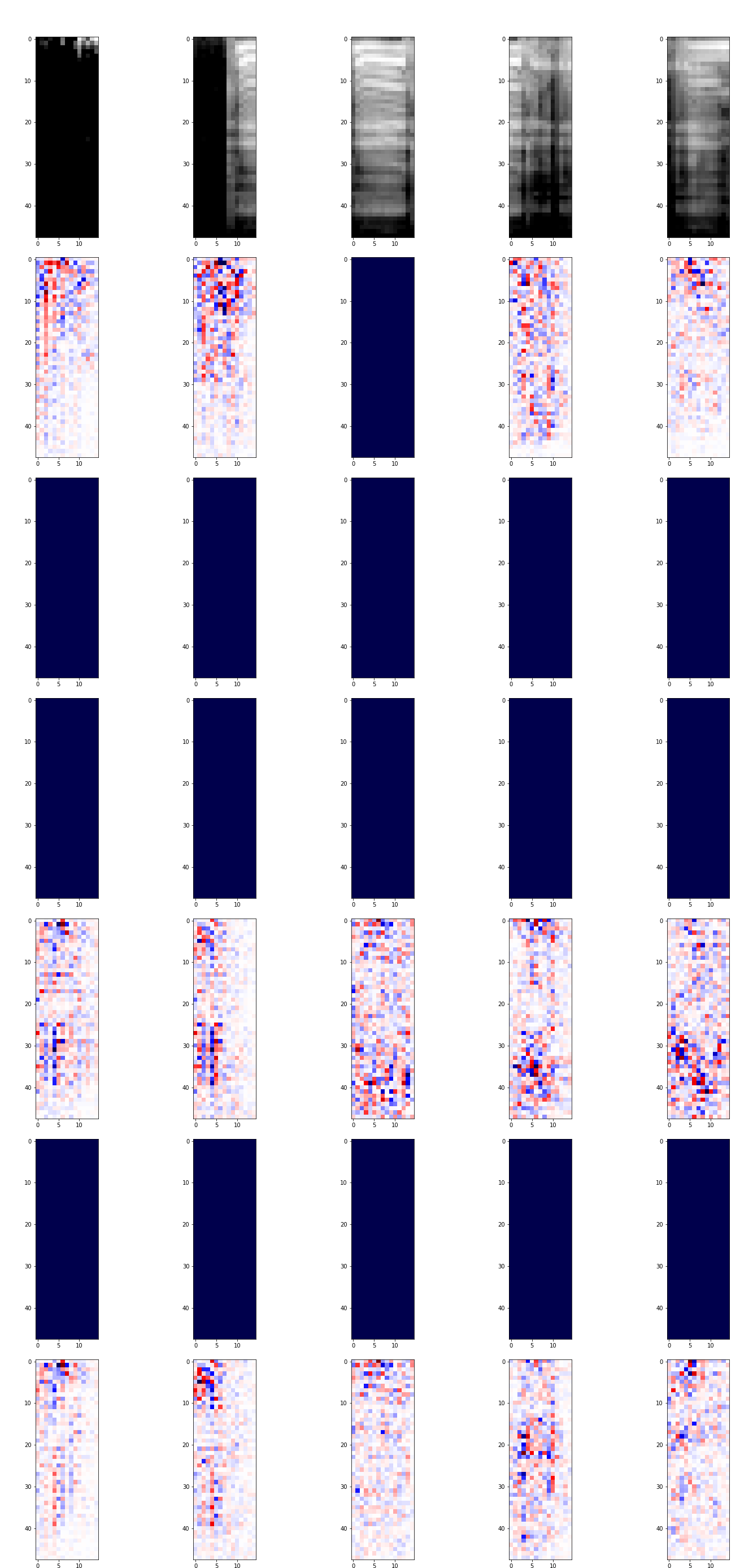}
    \caption{Explanations per segment (top row) w.r.t. different features (each one row). Only first 5 of 20 features depicted. Explanations were created with DeepLIFT.}
    \label{fig:few_segments}
\end{figure}

\begin{figure}[h!]
    \centering
    \includegraphics[width=0.35\textwidth]{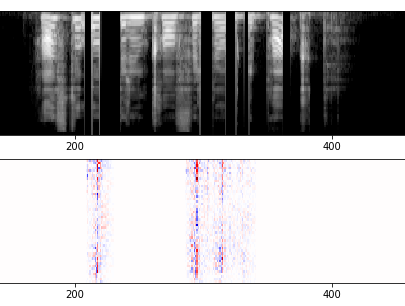}
    \caption{Spectrogram of an exemplary voice sample (top). Averaged explanations over all segments w.r.t. feature 17 (bottom). Explanation created with DeepLIFT.}
    \label{fig:avg_17}
\end{figure}

\begin{figure}[h!]
    \centering
    \includegraphics[width=0.45\textwidth]{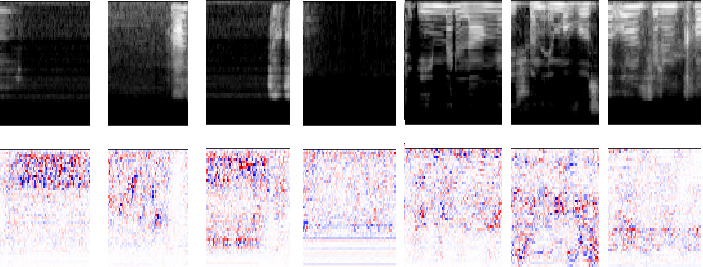}
    \caption{Cutouts of different spectrograms and different explanations, showing vertical orientation of features. From left to right [feature, vertical orientation, algorithm]: (13, top, DL), (0, top, DL), (11, top; bottom, DL), (19, top; bottom, DL), (5, center, IG), (8, center; bottom, IG), (3, bottom, DL).}
    \label{fig:vertical_regions}
\end{figure}

Integrated Gradients and DeepLIFT were used to check which parts of the input sample contribute to the implicitly learned features of NISQAs CNN. To identify those features the shape of the explanation is more important compared to the kind of explanation, namely if it's a positive or negative contribution. As exemplary depicted in Figure \ref{fig:few_segments}, an explanation for every segment with respect to every feature was created. It stands out that explanations for some features are empty while others have red-blue colored relevance maps. An empty explanation states that the respective feature is not present in the examined segment. For the explanations of both algorithms, it holds that some features are nearly always present, while some occur only on a few samples and some features are never detected in the entire dataset. The absence of some features can have two reasons. Either they occurred only on the much larger training dataset and are simply not present in this test set or during the training process these output nodes were discarded since only a smaller number of features was necessary to sufficiently learn the problem.\\
To get a better understanding of what the present features stand for, the explanations w.r.t. one feature for all segments of a sample were checked. An example for these averaged explanations can be seen in Figures \ref{fig:avg_17} and \ref{fig:avg_16,7}. While DL and IG use different approaches to produce their explanations the results came out qualitatively very similar. Before interpreting what the features represent, it is to mention that the algorithms in some cases produce explanations for empty segments. This behavior most likely results from a non-optimal choice of the baseline. This artifact leads to some horizontal stripes in the overall explanation due to the average-step.\\
The clearest interpretation has feature 17, for which an exemplary explanation is shown in Figure \ref{fig:avg_17}. Overall it is rarely active and seems to detect sharp disruptions in the samples. Since these kinds of disturbances lead to a strong decrease in the perceived quality, this feature strongly correlates with a low MOS value.
Another feature that is plain to see belongs to the 16th output node. As shown in Figure \ref{fig:avg_16,7} it corresponds to the absence of voice, nevertheless it is insensitive to the presence of noise. While features 16 and 17 are well distinguishable, the remaining features are either similar to others or not visually clearly interpretable. The features 0, 13 and 19 also show sensitivity for the regions without any speech but do not leave out the areas as clearly as feature 16 does. Feature 7 on the other hand reacts to the presence of speech and thus the opposite of the previously mentioned features, as also shown in Figure \ref{fig:avg_16,7}.
\newline
Besides the capability of detecting certain patterns of the input, some of the features seem to have a focus on a specific vertical region. Figure \ref{fig:vertical_regions} shows an overview of these different orientations. Despite occurring in explanations of most samples this vertical focus does not generalize to every sample. Therefore it remains to clarify whether the features are indeed specialized to certain frequencies or if this behavior is due to some hidden causes.
\begin{figure}[h!]
    \centering
    \includegraphics[width=0.47\textwidth]{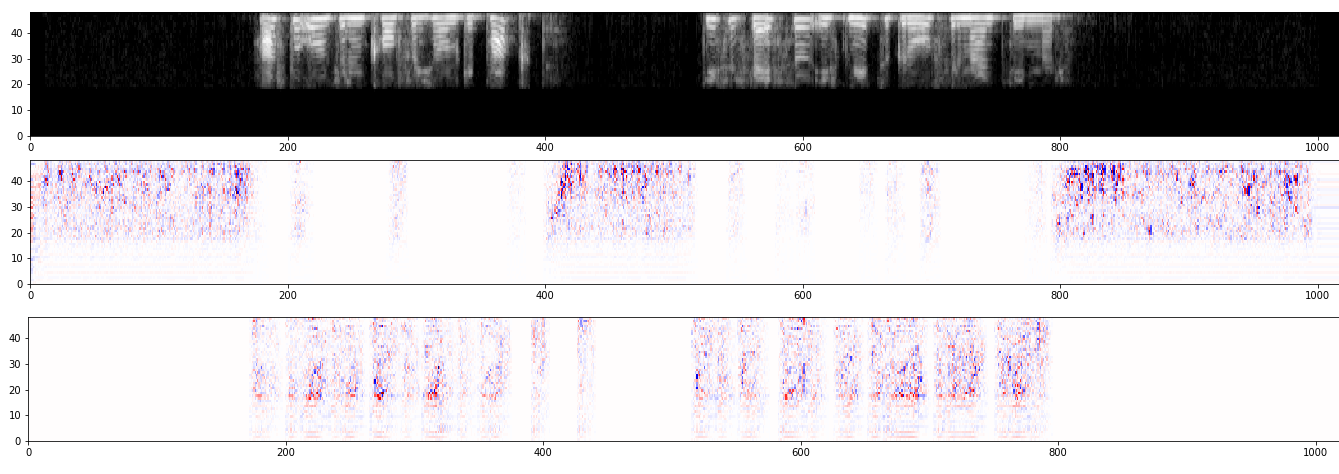}
    \caption{Spectrogram of an exemplary voice sample (top). Averaged explanation over all segments w.r.t. feature 16 (middle), feature 7 (bottom). Both explanations created with DeepLIFT.}
    \label{fig:avg_16,7}
\end{figure}

\begin{figure}[h!]
    \centering
    \includegraphics[width=0.3\textwidth]{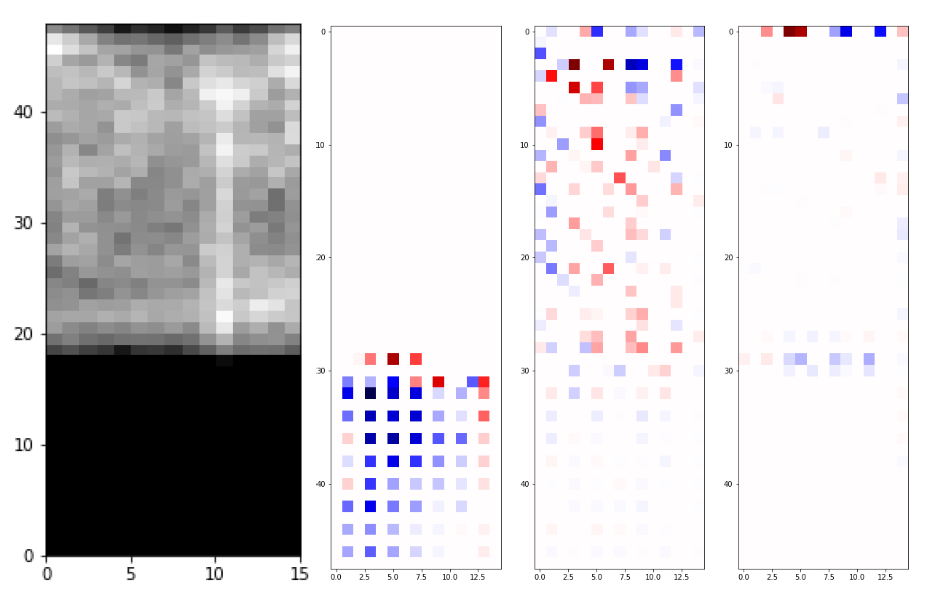}
    \caption{Segment of spectrogram (left) and explanations for different feature maps of the first convolutional layer w.r.t. feature 0. Explanations created with Conductance.}
    \label{fig:filter_all_region}
\end{figure}

\begin{figure}[h!]
    \centering
    \includegraphics[width=0.15\textwidth]{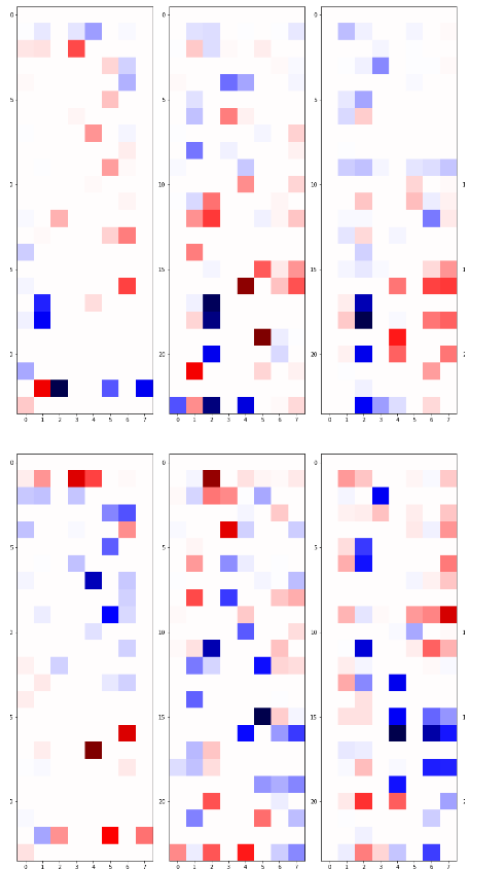}
    \caption{Explanations for one segment of different feature maps of the second convolutional layer w.r.t feature 8 (top row) and feature 11 (bottom row). Both features focus on the same input structure, but weight them in opposite ways.}
    \label{fig:filter_inverted}
\end{figure}

While IG and DL create explanations on the input layer Conductance examines the reasoning on the hidden convolutional layers of the CNN. In image classification it is well-known, that the first few layers are responsible to detect simple features, such as edges, enabling layers deeper in the network to find more complex shapes. This behavior could not be found in this experiment. However it is suspected that a similar process happens, but since the features do not have such a prominent shape as in the object detection case, they are just not visually detectable by a human observer. Nevertheless, some patterns in the convolutional layers are recognizable. The explanations of the first convolutional layer look quite similar to those of the input layer. The most notable behavior is, as shown in Figure \ref{fig:filter_all_region}, that different filters of the layer are responsible for different regions of the segment. This suggests that the model decomposes the problem into several subproblems.
%\newline
In general, the features with a similar explanation from IG and DL tend to have a similar explanation from Conductance. On the second convolutional layer, another interesting behavior appears. As depicted in Figure \ref{fig:filter_inverted} sometimes different features react to the same input pattern, but with an inverted explanation. For the following convolutional layers deeper in the network, no interpretable patterns could be identified. This is most likely due to the fact that their feature map size is rather small while the representational power lies within the high number of features maps. 
\section*{Conclusion}
The goal of this investigation was to better understand the reasoning of the speech quality assessment model NISQA. Using the simple Occlusion Sensitivity approach it was shown that NISQAs basic behavior works as intuitively expected. The more fine-grained techniques Integrated Gradients and DeepLIFT were able to identify groups of implicitly learned features that are used to estimate the speech quality. However, only the CNN component of NISQA was evaluated. To get a more complete idea of the reasoning process it would be beneficial to also include the LSTM part of NISQA to acquire an end-to-end explanation of the estimation. The used techniques also have a set of hyperparameters. Further fine-tuning may reveal additional details.

\bibliographystyle{ieeetr}
\bibliography{sample}

\end{document}